\documentclass[aip,pop,twocolumn, 10pt, floatfix]{revtex4-1} 
\usepackage{amsmath}
\usepackage{amssymb}
\usepackage{epsfig}

\usepackage{bm}

\newcommand{\vs}{\textit{vs.}\ }
\begin{document} 
\title{Comparing linear ion-temperature-gradient-driven mode stability of the National Compact Stellarator Experiment and a shaped tokamak} 

\renewcommand{\thefootnote}{\alph{footnote}}

\author{J. A. Baumgaertel}%
\affiliation{%
Los Alamos National Laboratory, Los Alamos, New Mexico 87544
}%
\author{G. W. Hammett}
\affiliation{%
 Princeton Plasma Physics Laboratory, Princeton, New Jersey 08543
}%
\author{D. R. Mikkelsen}
\affiliation{%
 Princeton Plasma Physics Laboratory, Princeton, New Jersey 08543
}%

\date{\today}

\begin{abstract} 
One metric for comparing confinement properties of different magnetic fusion energy configurations is the linear critical gradient of drift wave modes. The critical gradient scale length determines the ratio of the core to pedestal temperature when a plasma is limited to marginal stability in the plasma core.
The gyrokinetic turbulence code GS2 was used to calculate critical temperature gradients for the
linear, collisionless ion temperature gradient (ITG) mode in the National Compact Stellarator Experiment
(NCSX) and a prototypical shaped tokamak, based on the profiles of a JET H-mode shot and thencsx stronger shaping of ARIES-AT. While a concern was that the narrow cross section of NCSX at some toroidal locations would result in steep gradients that drive instabilities more easily, it is found that other stabilizing effects of the stellarator configuration offset this so that the normalized critical gradients for NCSX are competitive with or even better than for the tokamak. For the adiabatic ITG mode,
NCSX and the tokamak had similar adiabatic ITG mode critical gradients, though beyond marginal stability, NCSX had larger growth
rates. However, for the kinetic ITG mode, NCSX had a higher critical gradient and lower growth rates until $
a/L_T\approx1.5\ a/L_{T,crit}$, when it surpassed the tokamak's. A discussion of the results presented with
respect to $a/L_T$ vs $R/L_T$ is included.
\end{abstract}
\maketitle

\section{Introduction}

Two of the main magnetic fusion energy designs are the axisymmetric tokamak and the non-axisymmetric stellarator. Tokamaks have seen significant heat loss due to turbulence,\cite{liewer_measurements_1985} while stellarator losses have traditionally been dominated by their larger neoclassical transport. Studying turbulent transport in stellarators is increasingly important, however, as modern stellarator designs (such as Wendelstein 7-AS (W7-AS),\cite{sapper_stellarator_1990} Wendelstein 7-X (W7-X),\cite{beidler_physics_1990,grieger_physics_1992} the National Compact Stellarator Experiment (NCSX),\cite{zarnstorff_physics_2001} the Large Helical Device (LHD) \cite{yamada_configuration_2001}, and the Helically Symmetric Experiment (HSX) \cite{gerhardt_experimental_2005,canik_experimental_2007,talmadge_experimental_2008}) have shown or are designed to have improved neoclassical confinement and stability properties. Therefore, turbulence could be increasingly relevant in stellarator experiments.  Several gyrokinetic studies of drift-wave-driven turbulence in stellarator geometry have been done\cite{rewoldt_electromagnetic_1982,rewoldt_collisional_1987,rewoldt_drift_1999,xanthopoulos_nonlinear_2007, guttenfelder_effect_2008, xanthopoulos_gyrokinetic_2007, watanabe_gyrokinetic_2007,nunami_gyrokinetic_2010, baumgaertel_simulating_2011, baumgaertel_gyrokinetic_2012, boozer_stellarators_2008, rewoldt_comparison_2005, nunami_gyrokinetic_2012, yamagishi_particle_2007, watanabe_reduction_2008,watanabe_effects_2011, xanthopoulos_zonal_2011} with a variety of gyrokinetic codes, such as GS2,\cite{dorland_electron_2000} GENE,\cite{jenko_electron_2000, xanthopoulos_nonlinear_2007} GKV-X,\cite{watanabe_gyrokinetic_2007,nunami_gyrokinetic_2010} and FULL.\cite{rewoldt_electromagnetic_1982} These codes have all been linearly benchmarked against each other for non-axisymmetric geometries.\cite{baumgaertel_simulating_2011,baumgaertel_gyrokinetic_2012} Progress has even been made in optimizing stellarator designs to have reduced turbulent transport.\cite{mynick_optimizing_2010}

Besides comparing good stellarator configurations (as was done in Refs. \onlinecite{mynick_optimizing_2010,rewoldt_comparison_2005, baumgaertel_gyrokinetic_2012}, among others), one would like to compare stellarator confinement with that of tokamaks. The relative benefits of each device are important to consider when designing the next generation of experiments. A few previous comparison studies have been done, such as those in Refs. \onlinecite{boozer_stellarators_2008, rewoldt_comparison_2005}. Here, the gyrokinetic turbulence code GS2\cite{dorland_electron_2000} is used to compare microinstability of the electrostatic adiabatic ion temperature gradient (ITG) and the electrostatic collisionless kinetic ITG modes in the quasi-axisymmetric National Compact Stellarator Experiment (NCSX) design to that of a highly-elongated tokamak. Because this tokamak and NCSX geometry differs so significantly, it is hard to pinpoint what parameter has the greatest effect, but overall effects will be examined.

ITG mode-driven turbulence has been connected experimentally to measured heat losses in both tokamaks (e.g. Ref. \onlinecite{gruber_overview_2001}) and the LHD.\cite{nunami_gyrokinetic_2012} There is much variability in stellarator designs, and it is unclear without more study which modes will dominate in each. Preliminarily,  Ref. \onlinecite{mynick_reducing_2011} suggests that ITG transport in NCSX may be larger than that of ETG (see Figs. 8-9 of that paper). While only the ITG mode thresholds are compared in this paper, further study could show that other modes dominate in this case and in other devices.
 
In Section \ref{sec:ncsxtok}, a simple comparison metric is defined for use in this paper. Sections \ref{sec:miller}-\ref{sec:ncsx} describe the tokamak and stellarator configurations: a Miller equilibrium for the highly-elongated tokamak, and a numerical equilibrium for NCSX. Next, growth rates and critical temperature gradients are compared for the ITG mode in Sections \ref{sec:adie}-\ref{sec:kine}. Finally, the study is concluded in Section \ref{sec:concl}.

\section{NCSX  \vs a Shaped Tokamak}\label{sec:ncsxtok}

To understand the trade-offs between stellarator and axisymmetric geometry and their confinement capabilities, designs can be compared computationally. One metric of confinement quality is the ratio of the core temperature to the pedestal temperature, $T_0/T_{ped}$, as fusion reactors need very high core temperatures, and high core temperature implies good confinement.  This ratio is related to the critical temperature gradients. If $-\partial T/\partial r\approx T/L_{T,crit}$, temperature-gradient-driven instabilities are marginally stable--a reasonable assumption in a reactor plasma, as temperatures inside the pedestal will be so high that profile stiffness will ensure that gradients are close to marginal stability. This is demonstrated by Fig. 3 of Ref. \onlinecite{staebler_predicted_2006}, which shows that fusion power (and thus the temperature profile) depends primarily on the pedestal temperature and not the beam power, for the case of balanced beams.

At marginal stability (assuming $1/L_{T,crit}$ is independent of minor radius), 

\begin{equation}
T(r)=T_0e^{-r/L_{T,crit}}
\end{equation}
The minimum temperature is at the edge where $r$ is maximum, $r_{max}=a$, where $a$ is the minor radius. In this simplified story, another approximation will be made, that $T_{ped}$ occurs at $r=a$. So, $T(a)=T_{ped}=T_0e^{-a/L_{T,crit}}$. Therefore, the core temperature's dependence on the critical temperature gradient for typical tokamak values\cite{dimits_comparisons_2000, jenko_critical_2001, belli_effects_2008}  of $a/R\approx1/3.5$ and $R/L_{T,crit}\approx5$ is
 
\begin{equation}
\begin{array}{lcl}
T_0/T_{ped} &=& e^{a/L_{T,crit}} \\
&=&e^{(a/R)(R/L_{T,crit})} \\
&\approx & e^{(1/3.5)5}\approx4.2,
\end{array}
\label{eqn:coretemp}
\end{equation}

One wants to maximize the core temperature, $T_0$. $T_{ped}$ is set by non-transport mechanisms and cannot be arbitrarily high, leaving $a/L_{T,crit}$ as the important parameter in equation \ref{eqn:coretemp}. If an alternative fusion device design could increase $a/L_{T,crit}$ by just $30\%$, this would increase the central temperature by $50\%$, and more than double the fusion power. (A caveat--this is a simple estimate, and does not take into account the fact that MHD stability changes with higher pressure peakedness.) In this paper, the critical ion temperature gradients are compared for NCSX and a strongly-shaped tokamak design. The stated stellarator minor and major radii are the average values.

\subsection{Miller equilibrium for this tokamak}\label{sec:miller}
NCSX runs were compared to a potential high-elongation tokamak based on a composite of ARIES-AT \cite{najmabadi_aries-at_2006} and JET H-mode shot $\#$52979. It is well known that tokamak performance improves at high elongation and triangularity \cite{lazarus_higher_1991,strait_stability_1994,gates_high_2003} (in large part because this leads to more plasma current at fixed $q$), so when designing future tokamaks, one would like to use the highest possible values of elongation and triangularity, though elongation is limited by vertical stability control if it becomes too large. Some initial studies of shaping effects with GS2 were carried out in Ref. \onlinecite{belli_effects_2008}, using a range of shapes scaled from the particular JET shot $\#$52979 (available in the ITER profile database \cite{roach_public_2008}). This JET shot, described in more detail in Refs. \onlinecite{valovic_long_2002, ongena_towards_2004}, was chosen as a representative H-mode plasma that has been studied in detail by gyrokinetics codes before. Parameters for this paper were chosen from shot $\#$52979, but the shaping parameters were scaled to the higher levels achievable in tokamaks. These values of edge elongation and triangularity are from ARIES-AT, which generally tried to maximize these parameters subject to engineering and vertical stability constraints. This JET shot has a conventional $q$ profile, while the ARIES-AT design study assumed that a reversed shear scenario can be stably maintained in steady state. The composite tokamak of this paper has a conventional $q$ profile. The Miller equilibrium \cite{miller_noncircular_1998} for this prototypical or generic strongly-shaped tokamak was set up in the following way.

The shaping study in Ref. \onlinecite{belli_effects_2008} chose to focus on the radius $r/a=0.8$ in order to be fairly near the plasma edge where shaping effects are stronger, but not too far out because because gyrokinetic codes often do not compare well with experiments near the edge (perhaps because edge turbulence is driven by mechanisms other than ITG/TEM modes that require higher resolution than usual or additional effects that are not included in present gyrokinetic codes). Therefore, the study in this section uses $r/a=0.8$.

Just inside the separatrix at the $95\%$ poloidal flux surface, the JET shot had an elongation of $\kappa_{95}=1.73$ and triangularity of $\delta_{95}=0.46$, while the core elongation is $\kappa_{core}\approx1.3$. At the radius of interest, $r/a=0.8$, $R/a=3.42$, $\kappa_{0.8}=1.46$, $\kappa_{0.8}'=0.57$, $\delta_{0.8}=0.19$, $\delta_{0.8}'=0.60$, and the Shafranov shift is $\partial R_{0}/\partial r=-0.14$. Finally, the safety factor $q_{0.8}=2.03$ and magnetic shear $\hat{s}=1.62$. 

Keeping JET's Shafranov shift, $q$, and $\hat{s}$, a modified ARIES-AT case was created using its $\kappa_{95}=2.08$ and $\delta_{95}=0.76$. Assuming, from the tokamak shaping studies, that $\kappa'\propto(\kappa_{95}-\kappa_{core})$ and $\delta,\delta'\propto\delta_{95}$:
\begin{equation}
\kappa_{0.8}^{tok}=\kappa_{core}^{JET}+(\kappa_{0.8,JET}-\kappa_{core}^{JET})\frac{(\kappa_{95,tok}-\kappa_{core}^{JET})}{(\kappa_{95,JET}-\kappa_{core}^{JET})}=1.59
\end{equation}

\begin{equation}
 \kappa_{0.8}^{tok'}=\kappa_{0.8}^{JET'}\frac{(\kappa_{95,tok}-\kappa_{core}^{JET})}{(\kappa_{95,JET}-\kappa_{core}^{JET})}=1.03
  \end{equation}
  
\begin{equation}
 \delta_{0.8}^{tok}=\delta_{0.8}^{JET}\frac{\delta_{95}^{tok}}{\delta_{95}^{JET}}=0.31
 \end{equation}
 
\begin{equation}
\delta_{0.8}^{tok'}=\delta_{0.8}^{JET'}\frac{\delta_{95}^{tok}}{\delta_{95}^{JET}}=0.99
 \end{equation}

\begin{figure} \centering
\includegraphics[scale=1.0]{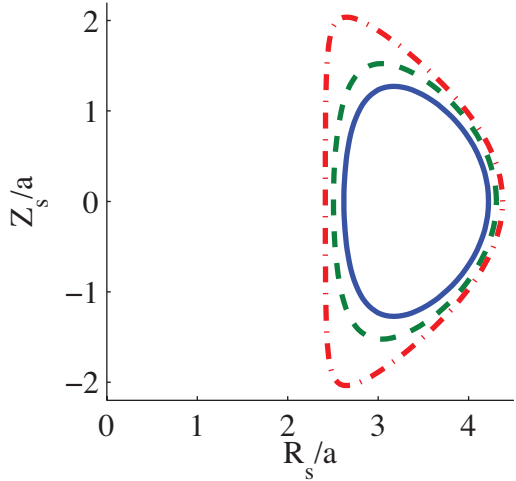}
\caption{Illustration of flux surface shapes for a prototype strongly-shaped tokamak at r/a = 0.8 (blue solid line), 0.9 (green dashed line), and 0.98 (red dash-dot line). (color online)}%
\label{fig:ariesFluxSurfs}
\end{figure}

Representative flux surfaces for this prototype strongly-shaped tokamak are shown in Figure \ref{fig:ariesFluxSurfs}.

\subsection{NCSX geometry}\label{sec:ncsx}

Refs. \onlinecite{baumgaertel_simulating_2011,baumgaertel_simulating_2012} describe how non-axisymmetric geometry input is created for GS2. The coordinate system of the flux-tube code GS2 includes the radial coordinate, $\rho=\sqrt{s}$ ($s\approx (r/a)^2$ is the normalized toroidal flux), the coordinate aligned to the field line, $\theta$, and the angle that selects a flux tube, $\alpha=\zeta-q(\theta-\theta_0)$ (where $\zeta$ and $\theta$ are Boozer toroidal and Boozer poloidal coordinates and $\theta_0$ is the ballooning parameter). 

The GS2 documentation\cite{barnes_trinity:_2009} defines geometrical quantities in terms of a parameter $d\Psi_N/d\rho$,  where $\rho$ is the radial coordinate and $\Psi_N$ is the normalized poloidal flux. Geometrical quantities in this paper follow GS2 notation and include $d\Psi_N/d\rho$. For more information, see Refs. \onlinecite{baumgaertel_simulating_2011,baumgaertel_simulating_2012}.

\begin{figure} \centering
\includegraphics[scale=1.0]{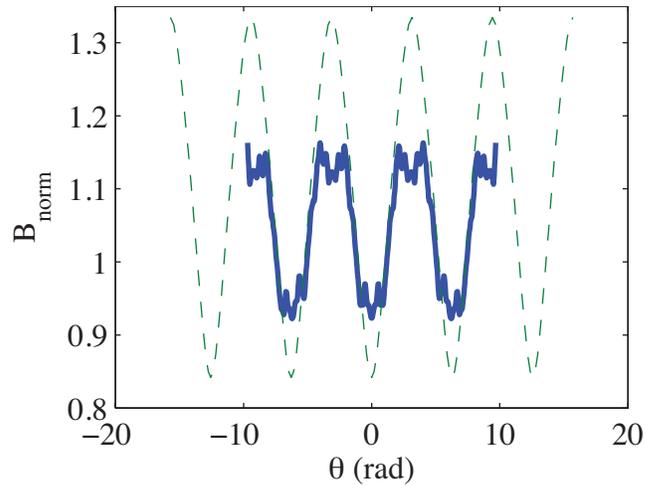}
\caption{The NCSX (blue solid line) and tokamak (green dashed line) equilibria: normalized $|B|$ \vs $\theta$. (color online)}
\label{fig:ncsxs30BMAG}
\end{figure}

\begin{figure} \centering
\includegraphics[scale=1.0]{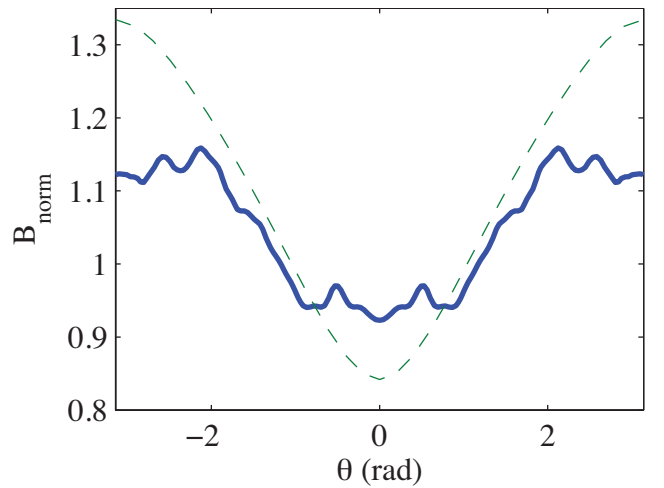}
\caption{The NCSX (blue solid line) and tokamak (green dashed line) equilibria: normalized $|B|$ \vs $\theta$, showing a close-up around $\theta=0$. (color online)}
\label{fig:ncsxs30BMAGz}
\end{figure}

\begin{figure} \centering
\includegraphics[scale=1.0]{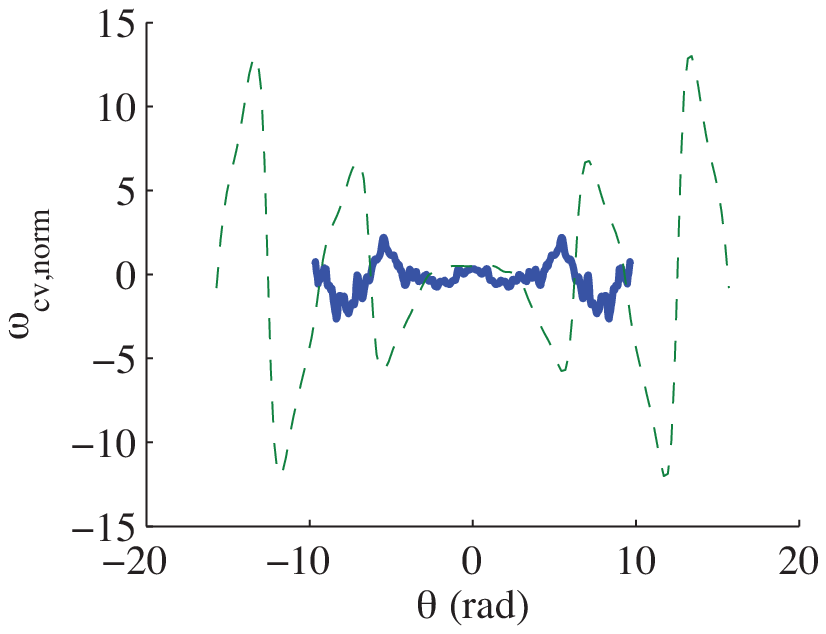}
\caption{The NCSX (blue solid line) and tokamak (green dashed line)  the curvature drift frequency ($\omega_{cv,norm}=(2a^2/B_N)(d\Psi_N/d\rho)(k_\perp/n)\cdot \mathbf{b}\times [\mathbf{b}\cdot\nabla \mathbf{b}]$) along $\theta$. (color online)}
\label{fig:ncsxs30DRIFTS}
\end{figure}

\begin{figure} \centering
\includegraphics[scale=1.0]{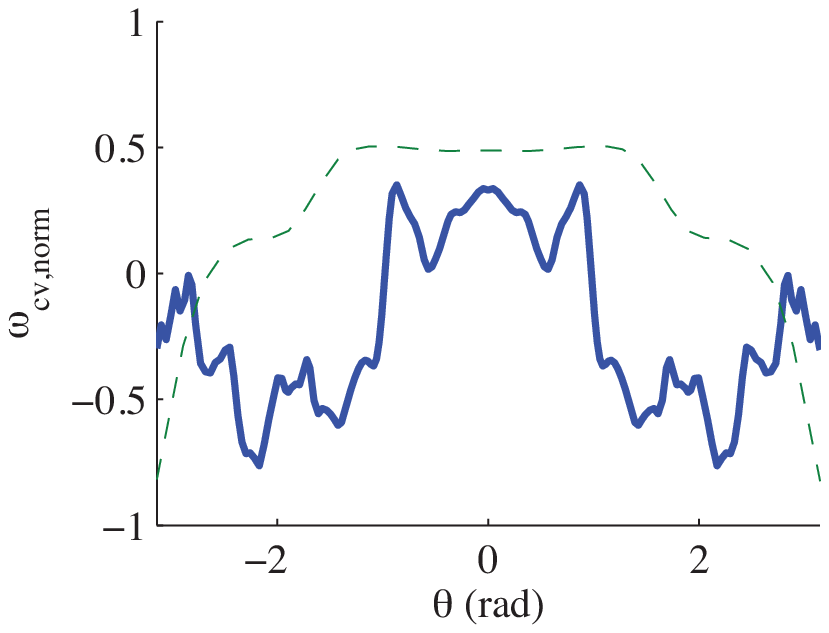}
\caption{The NCSX (blue solid line) and tokamak (green dashed line)  equilibria: the curvature drift frequency ($\omega_{cv,norm}=(2a^2/B_N)(d\Psi_N/d\rho)(k_\perp/n)\cdot \mathbf{b}\times [\mathbf{b}\cdot\nabla \mathbf{b}]$) along $\theta$, showing a close-up around $\theta=0$. (color online)}
\label{fig:ncsxs30DRIFTSz}
\end{figure}

\begin{figure} \centering
\includegraphics[scale=1.0]{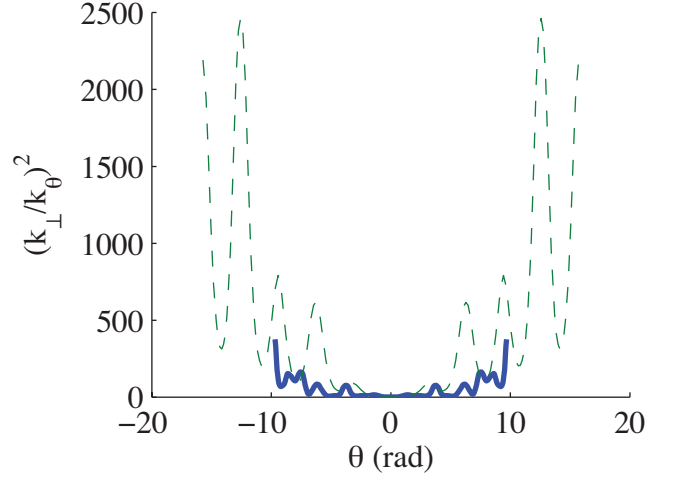}
\caption{The NCSX (blue solid line) and tokamak (green dashed line)  equilibria: $\left(\frac{k_\perp}{k_\theta}\right)^2$ \vs $\theta$. (color online)}
\label{fig:ncsxs30KPERP}
\end{figure}

\begin{figure} \centering
\includegraphics[scale=1.0]{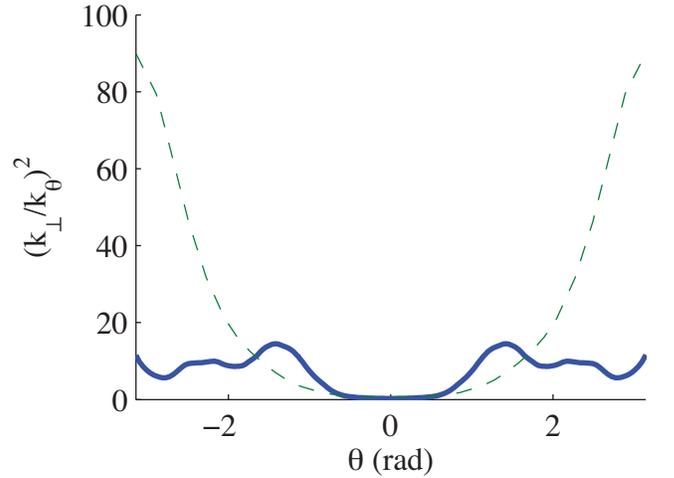}
\caption{The NCSX (blue solid line) and tokamak (green dashed line) equilibria: $\left(\frac{k_\perp}{k_\theta}\right)^2$ \vs $\theta$, showing a close-up around $\theta=0$. (color online)}\label{fig:ncsxs30KPERPz}
\end{figure}

The following figures show the magnitude of the magnetic field (Figs. \ref{fig:ncsxs30BMAG}-\ref{fig:ncsxs30BMAGz}), curvature drift (Figs. \ref{fig:ncsxs30DRIFTS}-\ref{fig:ncsxs30DRIFTSz}), and $(|k_\perp|/k_\theta)^2$ (Figs. \ref{fig:ncsxs30KPERP}-\ref{fig:ncsxs30KPERPz}) for both the tokamak and NCSX field lines, for the entire domain and a close-up around $\theta=0$. See Table \ref{tab:NCSXparams} for a complete list of geometrical quantities and their values.

\begin{table}
\centering
\begin{tabular}{|c|c|}
\hline 
 Parameter & Value \tabularnewline
\hline
$r/a$ & $0.8$\tabularnewline 
\hline
$s \approx \left(\langle r/a \rangle \right)^2$ & $0.64$\tabularnewline
\hline 
$\alpha=\zeta-q\theta$  & $0$\tabularnewline
\hline 
$\theta_{0}$  & $0$\tabularnewline
\hline 
$q_s$ & $1.70$ \tabularnewline
\hline 
$\hat{s}$& $0.835$\tabularnewline
\hline 
$\langle\beta\rangle$ & $0.0\%$\tabularnewline
\hline 
$R$ & $\approx4.7a_{N}\approx1.5m$ \tabularnewline
\hline 
$a_{N}$ & $\approx0.322m$\tabularnewline
\hline
$B_{a}=\langle B \rangle$ & $1.58 T$ \ \tabularnewline
\hline
\end{tabular}

\caption{Geometry values for the NCSX equilibrium.}
\label{tab:NCSXparams}
\end{table}

\begin{figure} \centering
\includegraphics[scale=1.0]{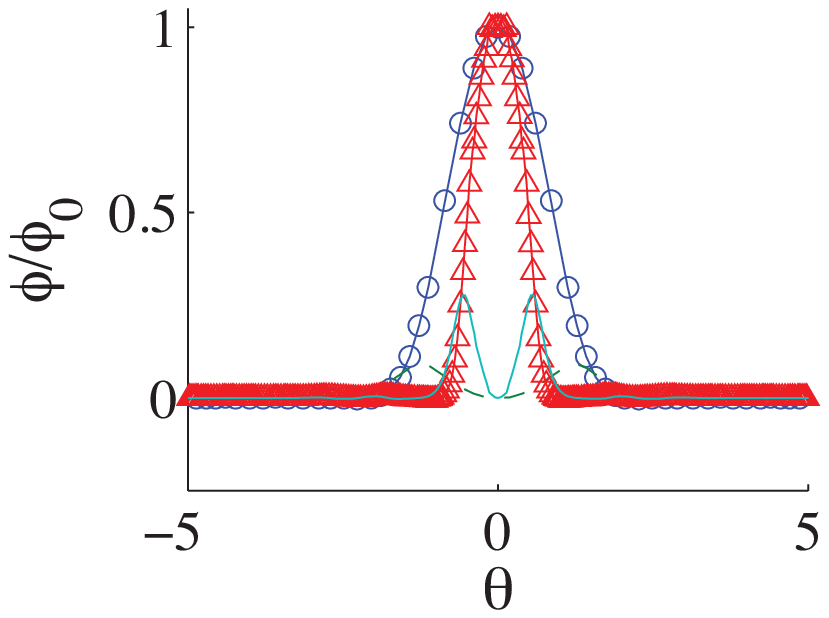}
\caption{Comparing electrostatic eigenfunctions for NCSX ($Re(\phi)$: red triangles and $Im(\phi)$: light blue solid line) and tokamak  ($Re(\phi)$: blue circles and $Im(\phi)$: green dashed line), for an adiabatic ITG mode with $a/L_T=3,a/L_n=0$. For ARIES, $k_y\rho_i=0.55$, and for NCSX, $k_y\rho_i=1.0$. (color online)}\label{fig:ncsxAries_efcns}
\end{figure}

Notice that the bad (positive) curvature regions of NCSX are much more localized than the tokamak case. Coupled with the much stronger local magnetic shear (responsible for the sharp peaks in $k_\perp\propto \hat{s}$, Fig. \ref{fig:ncsxs30KPERP}), this explains why NCSX's electrostatic potential eigenfunctions are also more localized than the tokamak's. An example is shown in Figure \ref{fig:ncsxAries_efcns}. These traits could predict better transport properties for NCSX.

\subsection{ITG mode with adiabatic electrons}\label{sec:adie}

\begin{figure} \centering
\includegraphics[scale=1.0]{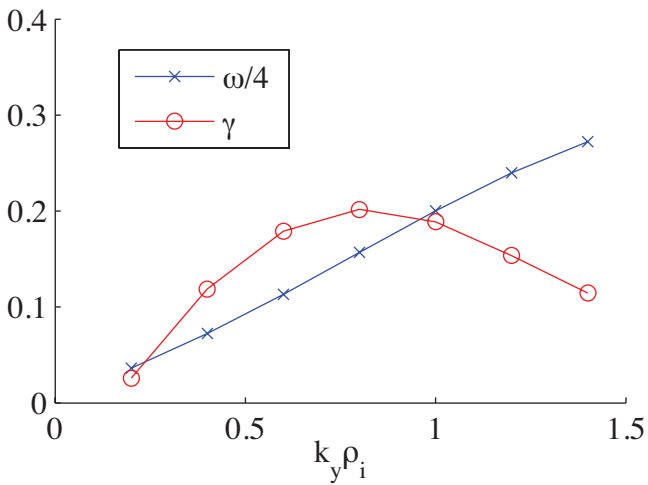}
\caption{NCSX growth rate spectrum for the adiabatic ITG mode with $a/L_T=3,a/L_n=0$. (color online)}\label{fig:ncsxAEKYRHO}
\end{figure}

\begin{figure} \centering
\includegraphics[scale=1.0]{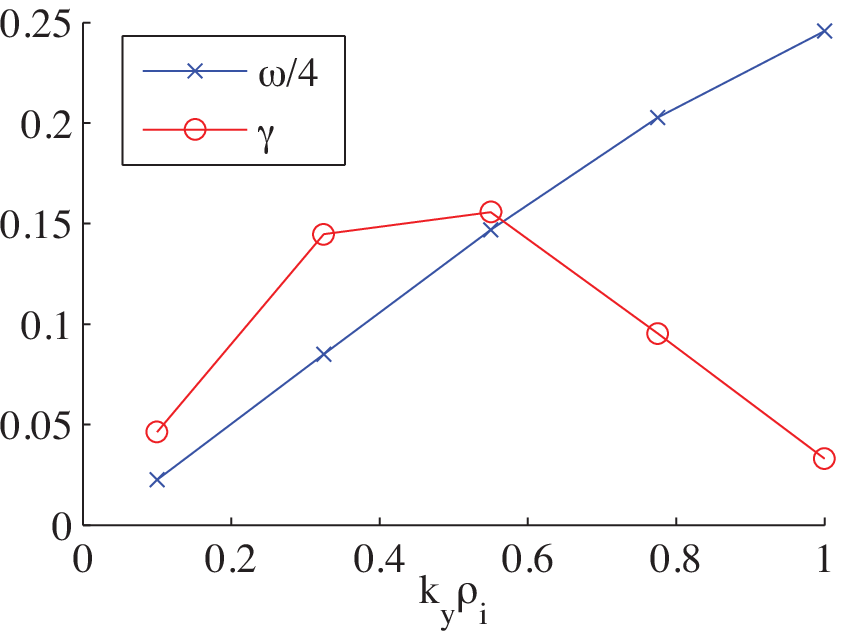}
\caption{Tokamak growth rate spectrum for the adiabatic ITG mode with $a/L_T=3,a/L_n=0$. (color online)}\label{fig:AriesAEKYRHO}
\end{figure}

\begin{figure} \centering
\includegraphics[scale=1.0]{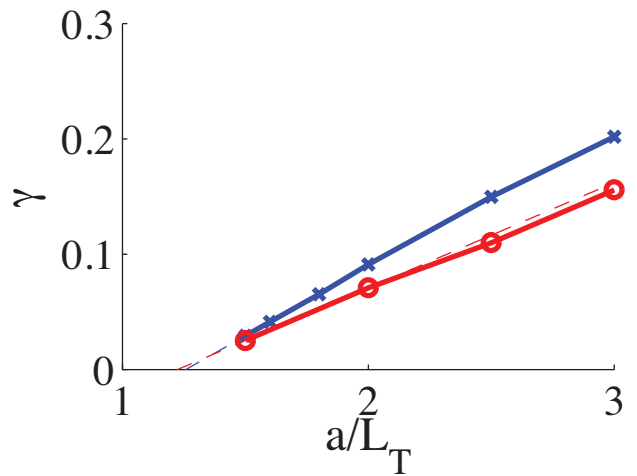}
\caption{NCSX (blue crosses) and ARIES-AT-like tokamak (red circles) adiabatic ITG mode growth rate dependence on temperature gradient. Fits obtained through piecewise linear interpolation on the lowest half of the growth rate curve. (color online)}\label{fig:ncsxAriesAE}
\end{figure}

For the initial study, the ITG mode with adiabatic electrons growth rates and their dependence on temperature gradient were compared. Figure \ref{fig:ncsxAEKYRHO}-\ref{fig:AriesAEKYRHO} show typical growth rate spectra for NCSX and this tokamak. In Figure \ref{fig:ncsxAriesAE}, the growth rate at each $a/L_T$ ($a/L_{Te}=a/L_{Ti}$) was the highest in the range $k_y\rho_i\in[0.2,1.4]$ for NCSX and $k_y\rho_i\in[0.1, 1.0]$ for the tokamak. These ranges were wide enough to capture the peak of the growth rate spectrum. Growth rates shown are normalized such that $(\gamma,\omega)=(\gamma_{physical},\omega_{physical})(a/v_{thi})$. The NCSX threshold is  $a/L_{T,crit}\approx1.26$ and the tokamak's is $a/L_{T,crit}\approx 1.22$. This difference is not very significant. However, soon after the threshold, the NCSX growth rates surpass those of the tokamak, indicating that for a given $a/L_T$, the adiabatic ITG mode is more unstable in NCSX than in the tokamak. This implies that the transport due to the adiabatic ITG mode would be stiffer, but the temperature gradients would still be expected to be very similar since they would be set by $a/L_{T,crit}$.


\subsection{ITG mode with kinetic electrons}\label{sec:kine}

\begin{figure} \centering
\includegraphics[scale=1.0]{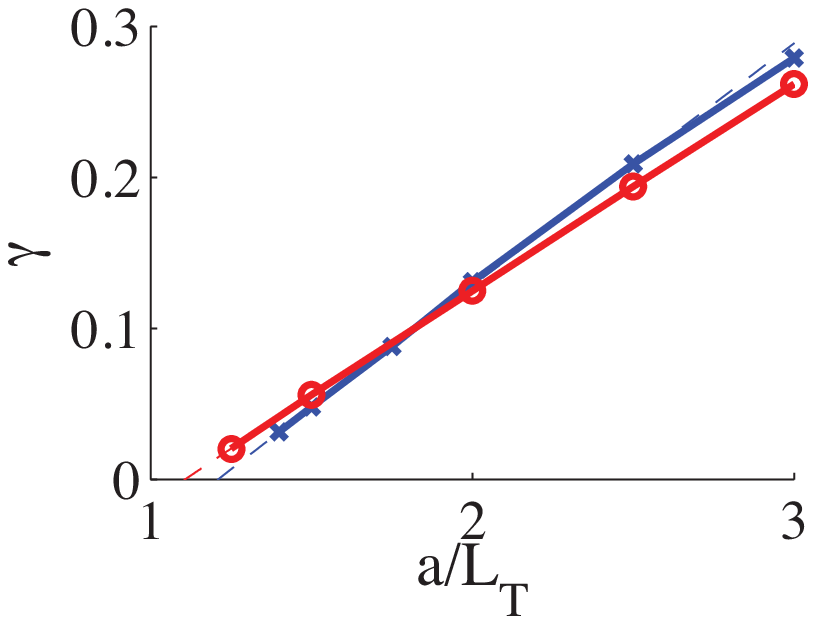}
\caption{Growth rates for an ITG mode with kinetic electrons as a function of temperature gradient for NCSX (blue crosses) and an ARIES-AT-like tokamak configuration (red circles). Fits obtained through piecewise linear interpolation on the lowest half of the growth rate curve. (color online)}
\label{fig:ncsxAries}
\end{figure}

\begin{figure} \centering
\includegraphics[scale=1.0]{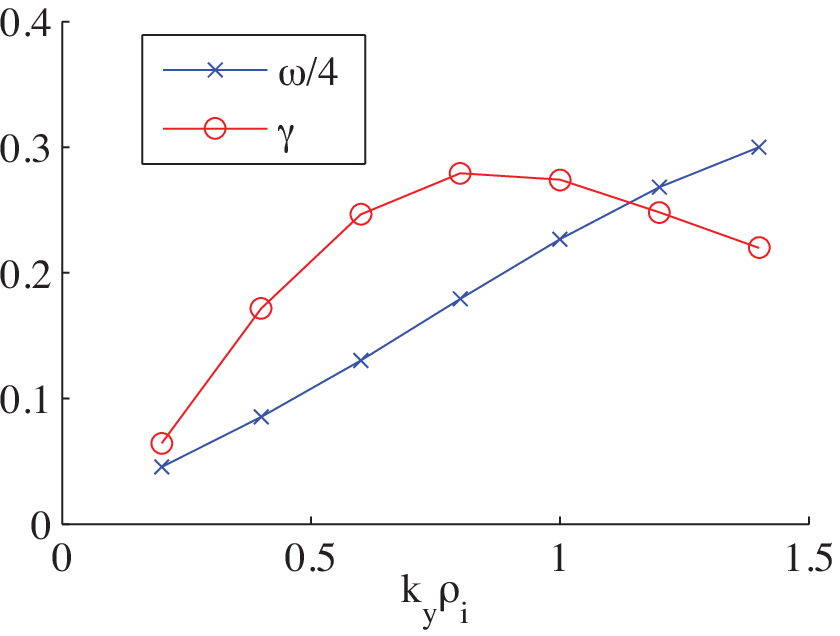}
\caption{NCSX growth rate spectrum for the kinetic ITG mode with $a/L_T=3,a/L_n=0$. (color online)}\label{fig:ncsxKEKYRHO}
\end{figure}

\begin{figure} \centering
\includegraphics[scale=1.0]{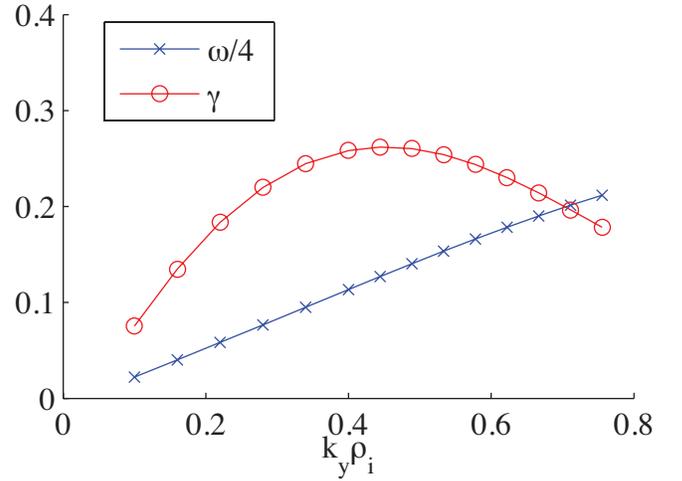}
\caption{Tokamak growth rate spectrum for the kinetic ITG mode with $a/L_T=3,a/L_n=0$. (color online)}\label{fig:AriesKEKYRHO}
\end{figure}

The threshold of the ITG mode with kinetic electrons (with $a/L_n=0$) for the tokamak was somewhat lower than that of NCSX, but the slope of the growth-rate curve is almost the same for both (Fig. \ref{fig:ncsxAries}). Similar to Section \ref{sec:adie}, growth rates shown were the highest on a spectrum of $k_y\rho_i\in[0.2,1.4]$ for NCSX and $k_y\rho_i\in[0.1,1.0]$ for the tokamak (see  Figure \ref{fig:ncsxKEKYRHO}-\ref{fig:AriesKEKYRHO} show typical growth rate spectra). With kinetic electrons, the growth rates for the ITG mode in NCSX increased over the adiabatic electron case (Fig. \ref{fig:ncsxAriesAE}), while the critical gradient lowered to $a/L_{T,crit}\approx1.21$.  The tokamak threshold decreased somewhat further, to $a/L_{T,crit}\approx1.11$. The slope of the NCSX line is somewhat steeper, and for $a/L_{T}\approx1.82$, the NCSX growth rates are larger than the tokamak growth rates.


The growth rate \vs $a/L_T$ plot in Figure \ref{fig:ncsxAries} shows improvement in the $a/L_T$ threshold of NCSX over this tokamak by about $10\%$. Based on the marginal stability logic in the beginning of Section \ref{sec:ncsxtok}, this corresponds to about $22\%$ more fusion power for a NCSX-based design relative to the tokamak (with the same edge temperature and density assumed for the two designs, and approximating the fusion power as scaling as $T^2$). Effects that might change this result include finite beta modifications to the equilibrium and the nonlinear Dimits shift,\cite{dimits_comparisons_2000, belli_effects_2008} which could increase each critical gradient, but the required nonlinear simulations are beyond the scope of this work. A Dimits shift has been reported for a stellarator,\cite{xanthopoulos_nonlinear_2007} and has been found in tokamak simulations.\cite{mikkelsen_dimits_2008, peterson_suppressing_2012}

This is much better than one might have initially guessed based on just the local value of $R/L_T$ in NCSX \vs a tokamak. While from equation \ref{eqn:coretemp} it is clear that $a/L_T$ is the relevant parameter for determining the core temperature, in the axisymmetric community, the threshold for the ITG instability is usually expressed in terms of $R/L_T$, which is often the key parameter numerically.  An instability threshold $R/L_{T,{ crit}}$ can be derived from the dispersion relation for a local ITG mode in the bad curvature region, ignoring the parallel dynamics.  In this limit, the critical instability parameter is the ratio of the temperature-gradient diamagnetic drift frequency ($\omega_{*T} \propto 1/L_{T_i}$) to the curvature drift frequency ($\omega_d \propto 1/R$). (Particles with different energies have different curvature drift velocities, which can result in Landau damping. This criterion essentially says that the drive from the temperature gradient must be strong enough to overcome this damping in order to drive instabilities.)

\begin{figure} \centering
\includegraphics[scale=1.0]{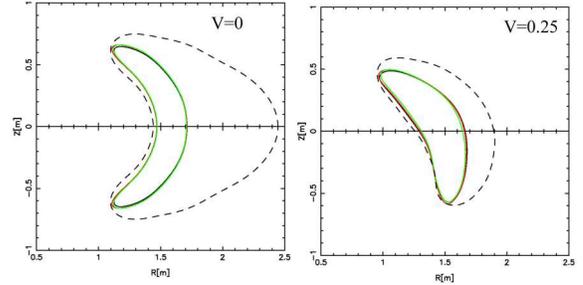}
\caption{Poloidal cross sections of NCSX for two toroidal angles. The dashed line is the location of the vacuum vessel and the solid lines are last closed flux surfaces for various $\iota$ profiles. Figure 2 of Ref. \onlinecite{pomphrey_flexibility_2002}; reprinted with permission. More information can be found in Ref. \onlinecite{pomphrey_ncsx_2007}.}\label{fig:cross}
\end{figure}

A concern could be that if an NCSX design is limited to the same local $R_{ loc}/L_{T,{ loc}}$ as in a tokamak, it would have a much lower $a/L_T$ (because at some toroidal locations, such as the left panel of Figure \ref{fig:cross}, the cross section of NCSX is very narrow, with a local plasma half-width $a_{loc} \approx a /2.58$), and thus would have much lower fusion power. Therefore, the local value of the logarithmic gradient, $1/L_{T,{ loc}} = |\nabla T|/T = (a/L_T)/a_{ loc}$ is much larger than the average $1/L_T$.  This is enhanced by the larger average aspect ratio $(R/a)_{ NCSX} = 4.7$ relative to the tokamak $(R/a)_{ tok}=3.42$, and is partially compensated by the fact that the local radius of curvature of the magnetic field, $R_{ loc} = |\hat{b}\cdot \nabla \hat{b}|^{-1} = 0.92 \mathrm{ m}$ (evaluated at the outer midplane of the $r/a=0.8$ flux surface in the left panel of Figure \ref{fig:cross}) is somewhat smaller than the average radius of curvature $R = 1.51 \mathrm{ m}$ in NCSX.  Considering these modifications, the local $R_{loc}/L_{T,{ loc}} = (R_{ loc}/R) (a/a_{ loc}) (R/a) (a/L_T) = 7.39\ a/L_T$ in NCSX, while $R/L_T = 3.42\ a/L_T$ for a tokamak.

\begin{figure} \centering
\includegraphics[scale=1.0]{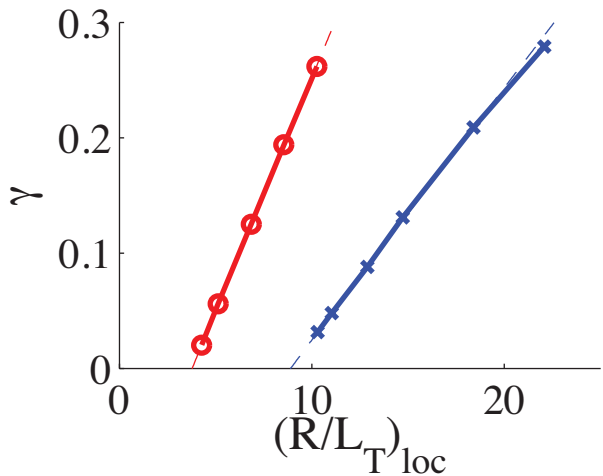}
\caption{Similar to Fig. \ref{fig:ncsxAries}, except the x-axis is normalized by the local magnetic field radius of curvature $R_{ loc}$, instead of $a$. This demonstrates that NCSX (blue crosses) performs much better than would be expected if the instability was the same at the same $(R/L_T)_{ loc}$, presumably indicating that additional stabilizing effects in the parallel dynamics are important in NCSX. Tokamak growth rates: red circles. (color online)}
\label{fig:ncsxAriesRLTloc}
\end{figure}

Restating this concern, one may have thought that if NCSX and a tokamak had the same normalized temperature gradient, $a/L_T$, the ITG modes would be much worse in NCSX due to a much higher $R_{ loc}/L_{T,{ loc}}$ than the tokamak.  In fact, Figure \ref{fig:ncsxAries} showed that NCSX has a somewhat higher critical gradient in terms of $a/L_T$, so the hypothesis that NCSX and a tokamak are similar when expressed in terms of $R_{ loc}/L_{T,{ loc}}$ must be incorrect. Indeed, this is strikingly illustrated in Figure \ref{fig:ncsxAriesRLTloc} (same data as Fig. \ref{fig:ncsxAries}, renormalized), which shows that NCSX has in fact much lower growth rates than a tokamak for the same $R_{ loc}/L_{T,{ loc}}$. This is probably because the parallel dynamics are in fact not negligible in NCSX. The eigenfunctions, as seen in Figure \ref{fig:ncsxAries_efcns}, are more localized along a field line in NCSX than in a tokamak, possibly through some combination of the stabilizing effects of a narrower bad-curvature region (i.e., a shorter connection length between good and bad curvature regions) and stronger local magnetic shear. These effects should be investigated more thoroughly in the future.

\section{Conclusion}\label{sec:concl}

Cross-configuration comparisons of plasma confinement are important to consider for the design of future fusion energy devices. As a simple case, the linear stability of the adiabatic and kinetic ITG modes was compared for NCSX and a tokamak equilibrium. This particular tokamak equilibrium is a composite of JET H-mode shot $\#$52979 and ARIES-AT. NCSX had a similar linear critical temperature gradient $a/L_{T,crit}$ to the tokamak case for ITG modes with adiabatic electrons, though its growth rates were higher than the tokamak's beyond marginal stability. However, for ITG modes with kinetic electrons, NCSX's critical gradient $a/L_{T}$ is approximately $9\%$ higher than the tokamak's, which would correspond to an approximately $22$\% increase in the fusion power for NCSX relative to the tokamak.  The growth rates in NCSX remained less than for the tokamak until $a/L_T \gtrsim 1.5 \ a/L_{T,crit}$. 

The parameter $a/L_{T,crit}$ is an important figure of merit because it characterizes the core to edge temperature ratio (if the plasma is near marginal stability as expected in typical hot reactor regimes).  While the parameter $R/L_{T,crit}$ is often a useful stability parameter in tokamak cases, it was found that stabilizing effects in the parallel dynamics in stellarators can make it a less relevant measure for stellarators. Upon rescaling the kinetic ITG mode data as a function of $R/L_T$, it was found that NCSX appears even more stable. 

Future work that should be done includes using GS2's nonlinear capabilities to compare heat fluxes for various fusion energy devices. Including more physical effects, such as non-zero density gradients and collisionalities, would create a clearer picture of their relative confinement properties. A future study could compare stellarators with tokamaks in various operating regimes that may potentially improve performance further, including reversed magnetic shear and hybrid low-shear scenarios. 

\section{Acknowledgments}

The authors wish to thank Neil Pomphrey for creating the NCSX equilibrium, Pavlos Xanthopoulos for the use of and assistance with GIST, and W. Dorland, M. A. Barnes, and W. Guttenfelder for their help with GS2. This work was supported by the U.S. Department of Energy through the SciDAC Center for the Study of Plasma Microturbulence, the Princeton Plasma Physics Laboratory under DOE Contract No. DE-AC02-09CH11466, and Los Alamos National Security, LLC under DOE Contract No. DE-AC52-06NA25396.
 
%

\end{document}